# High-Resolution TOF with RPCs

P. Fonte [1,2,*], V. Peskov[3]

1 – CERN-EP, Geneva, Switzerland
2 – LIP, Coimbra, Portugal
3 – Royal Institute of Technology, Stockholm, Sweden.

**Abstract**

In this work we describe some recent results concerning the application of Resistive Plate Chambers operated in avalanche mode at atmospheric pressure for high-resolution time-of-flight measurements.

A combination of multiple, mechanically accurate, thin gas gaps and state-of-the-art electronics yielded an overall (detector plus electronics) timing accuracy better than 50 ps $\sigma$ with a detection efficiency up to 99% for MIPs. Single gap chambers were also tested in order to clarify experimentally several aspects of the mode of operation of these detectors.

These results open perspectives of affordable and reliable high granularity large area TOF detectors, with an efficiency and time resolution comparable to the existing scintillator-based TOF technology but with a significantly, up to an order of magnitude, lower price per channel.

[*] Corresponding author: Paulo Fonte, LIP - Coimbra, Departamento de Física da Universidade de Coimbra, 3004-516 Coimbra, PORTUGAL.
tel: (+351) 239 833 465, fax: (+351) 239 822 358, email: fonte@lipc.fis.uc.pt

## 1. - Introduction

Heavy-ion collision physics at very high energies is emerging in many accelerator centres around the world (RHIC at BNL, SIS at GSI, and LHC-HI at CERN), emphasising the need for large area particle identification systems able to cope with high particle multiplicities. Time of Flight (TOF) sub-detectors, in particular, are foreseen for many such experiments (STAR, FOPI, ALICE), stimulating R&D on new, more cost effective, approaches to the timing of MIPs.

In this paper we describe some recent work done in the framework of the ALICE experiment on the development of Resistive Plate Chambers (RPCs) for TOF measurements (timing RPCs). The work included beam tests of single chambers and of a multichannel TOF prototype equipped with such chambers.

Some studies aimed to clarify experimentally the origin of the very good detection efficiency (99%) observed for MIPs in our timing RPCs will be also described.

## 2. - Experimental setup

Details on the mechanical construction, electronics, experimental setup and data analysis for the timing RPCs were already given elsewhere [1] and we will describe here only the few modifications introduced for the single cell tests.

Timing RPCs were made with glass [2] and aluminium electrodes forming a pair of double-gap chambers. The four gas gaps of 0.3 mm were accurately defined by glass optical-fiber spacers. The single gap RPCs were built essentially along the same lines, being a schematic drawing shown in Figure 1.

The signals were sensed by a custom made pre-amplifier [1], whose output was split in 3 identical channels via analogue buffers. One of the outputs was directly fed to a LeCroy 2249W ADC that measured a charge proportional to the total signal charge. Another output had the ion signal component (1 to 3 μs long) cancelled by forming the difference between the signal and it´s image delayed by 16 ns. A LeCroy 2249A ADC integrated the resulting short pulse, measuring a charge proportional only to the electron (fast) component of the signal. The third output was further amplified by a factor 10 and fed to a custom-made fixed threshold discriminator (typically set at a level equivalent to a total signal charge of 0.2 pC), followed by a LeCroy 2229 TDC with a 50 ps bin width.

The total charge was calibrated by injecting to the test input of the pre-amplifier a current pulse of an intensity and width similar to the ion current pulse from the chambers.

## 3. – Results

### *3.1 – Timing RPCs*

Timing RPCs in a single channel configuration have shown timing resolutions below 50 ps σ with an efficiency of 99% for MIPs [1]. A typical signal charge distribution is shown in Figure 2.

A 32-channel prototype equipped with similar chambers and suitable multichannel electronics has shown an average time resolution of 88 ps σ with a spread of 9 ps and an average efficiency of 98 % with a spread of 0.5 % [3]. The crosstalk between neighbouring channels generally did not exceed 1%.

*3.2 - Single gap chambers*

In Figure 3 we show the signal charge distributions measured at different applied voltages in single gas gaps of 0.1 and 0.3 mm filled with methane, isobutane or a “standard RPC mixture” containing $C_2H_2F_4$+10%$SF_6$+5%isobutane, also used for the timing RPCs.

In pure isobutane and in the “standard mixture”, for both gap sizes, the distributions show an extended flat region for the larger applied voltages. This is quite surprising because it can be shown theoretically[1] that the event by event variations in the position of the leading cluster (closer to the cathode) causes strong fluctuations in the final avalanche size, resulting in a charge distribution almost proportional to 1/Q (being Q the signal charge). The observed distribution is much more favourable for an efficient particle detection than the expected “1/Q” distribution, allowing for the excellent detection efficiency measured in chambers with four gaps (see also Figure 2).

In methane only a modest gas gain could be reached due to the onset of discharges, presumably caused by photon feedback. Indeed, while for the other gases all results were essentially independent of the cathode material (aluminium or glass), in methane larger gains could be reached with the glass cathode, presumably due to the smaller quantum efficiency of glass.

Even for a single 0.1 mm gas gap the detection efficiency was still around 45 % (isobutane). From the corresponding inefficiency figure and the gap length it can be calculated that the mean number of primary clusters per unit length must be at least 6 $mm^{-1}$ (at least one cluster must be produced for a particle to be detected). However, due to the exponential dependence of the final avalanche size on the cluster position, only a small region of the gas gap (closer to the cathode) will be sensitive to the ionising particles. Therefore, to explain the observed efficiency, the ionisation density must be a few times larger than the figure calculated above, being quite doubtful whether this is physically possible [4]. In fact there is some indication that a process other than gas ionisation may be also contributing to the observed detection efficiency (Figure 4).

Some further information can be obtained by plotting the average signal charge as a function of the applied field (Figure 5). There is a strongly sub-exponential growth of the average charge with the applied field, indicating the presence of a gas gain saturation effect.

Although similar effects have been observed in parallel geometry counters at low pressures [5] and wire counters, being generally attributed to space-charge effects, such strong saturation has never been, to the author’s knowledge, observed at atmospheric pressure in proportional-mode parallel geometry counters. This may be related to the fact that our detectors work in an E/p range (around 100 V/cm Torr) similar to those typically found in low-pressure parallel-plate counters or at the surface of wires in cylindrical counters and MWPCs.

Further evidence of a strong space charge effect is presented in Figure 6, were correlation plots between the fast (electron) signal charge and the total signal charge were drawn for the same experimental conditions studied above. Standard detector theory [5] shows that the ratio between these quantities should be independent of the avalanche size and equal to $(\alpha v)^{-1}$, where $\alpha$ is the First Townsend Coefficient and $v$ is the electron drift velocity. This situation is observed for the case of a 0.1 mm gap filled with methane. For all other cases the upward

1 See for instance the Appendix in ref. [6].

curving correlation is compatible with a space charge effect that would reduce the effective value of $a$ for the larger avalanches [5].

## 4. – Conclusions

Timing RPCs made with glass and metal electrodes, forming four of accurately spaced gas gaps of 0.3 mm, have reached time resolutions below 50 ps $\sigma$ with a detection efficiency of 99% for MIPs [1].

A 32-channel prototype equipped with such chambers has shown an average resolution of 88 ps $\sigma$ with a spread of 9 ps and an average efficiency of 98 % with a spread of 0.5 %. The crosstalk between neighbouring channels generally did not exceed 1% [3].

It was found that in single gas gaps the signal charge distribution departs strongly from the theoretically expected shape and that the gas amplification process seems to be strongly influenced by a space charge effect. This effect may be related to the unexpected charge distribution observed.

The relatively large detection efficiencies observed in single gas gaps (up to 45% for 0.1 mm gaps and up to 75% for 0.3 mm gaps) seem to be incompatible with a primary detection process based uniquely in the ionisation of the gas by the incoming particles.

These conclusions seem to be quite independent on the nature of the filling gas, applying both to the operation in pure isobutane and in a strongly electronegative mixture containing Freon and $SF_6$.

## 5. – Acknowledgements

The use of the instrumented beam line installed by the ALICE experiment for the tests in the T10 PS beam under the supervision of W. Klempt is acknowledged, as well as the kind support of the CERN EP/AIT group. The data acquisition infrastructure and the tracking system were implemented and managed by Paolo Martinengo.

We are also very grateful to P.G. Innocenti, W. Klempt, C. Lourenço, G. Paic, F. Piuz, R. Ribeiro and J. Schukraft for their comments and suggestions and for their interest on our work.

We benefited also from many discussions and from the accumulated experience of the members of the ALICE-TOF project and from the technical expertise of Dave Williams.

This work was partially supported by the FCT research contract CERN/FAE/1197/98. One of us (V. Peskov) acknowledges the financial support of LIP-Coimbra.

## 6. - References

**Figure captions**

Figure 1 - Structure of a single-gap detector cell.

Figure 2 – Comparison between the signal charge distribution observed in a timing RPC (histogram) with four gas gaps and the 4-fold self-convolution of the charge distribution from a single-gap chamber (solid line) measured in similar operating conditions. The overall good agreement between both distributions suggests that the charge distribution observed in the four-gaps chamber can be interpreted as the analog sum of four independent single gaps.

Figure 3 – Distribution of the signal charge in single gap chambers for several applied voltages, filling gases and gap widths. The innermost distributions correspond to the ADC pedestal and the peak close to 0 pC corresponds to the detector inefficiency. The efficiency figures were measured by the method described in [2].

Figure 4 – The extrapolation of the observed detection efficiency to a vanishing gas gap width suggests that some additional process, other than gas ionization, may be contributing to the detection efficiency.

Figure 5 – Average signal charge as a function of the applied electrical field, calculated from the data presented in Figure 3. The solid lines correspond to exponential functions fitted to the lower 3 points of each experimental series and extrapolated to the larger fields, evidencing the sub-exponential character of experimental data (gain saturation). The onset of gain saturation for the 0.1 mm gaps occurs at a charge level that is an order of magnitude smaller than for the 0.3 mm gaps.

Figure 6 – Correlation plots between the fast (electron) signal charge and the total signal charge. Standard detector theory shows that the ratio between these quantities should be a constant, being the observed upward curving correlation compatible with a space charge effect.

**Figures**

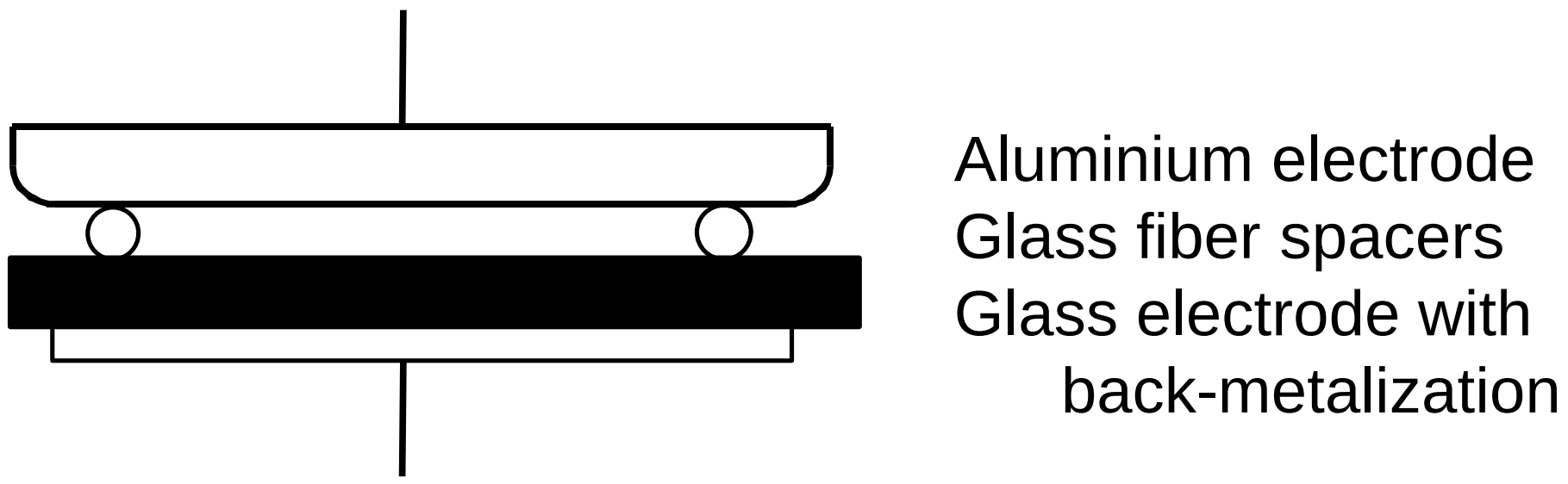

Figure 1 - Structure of a single-gap detector cell.

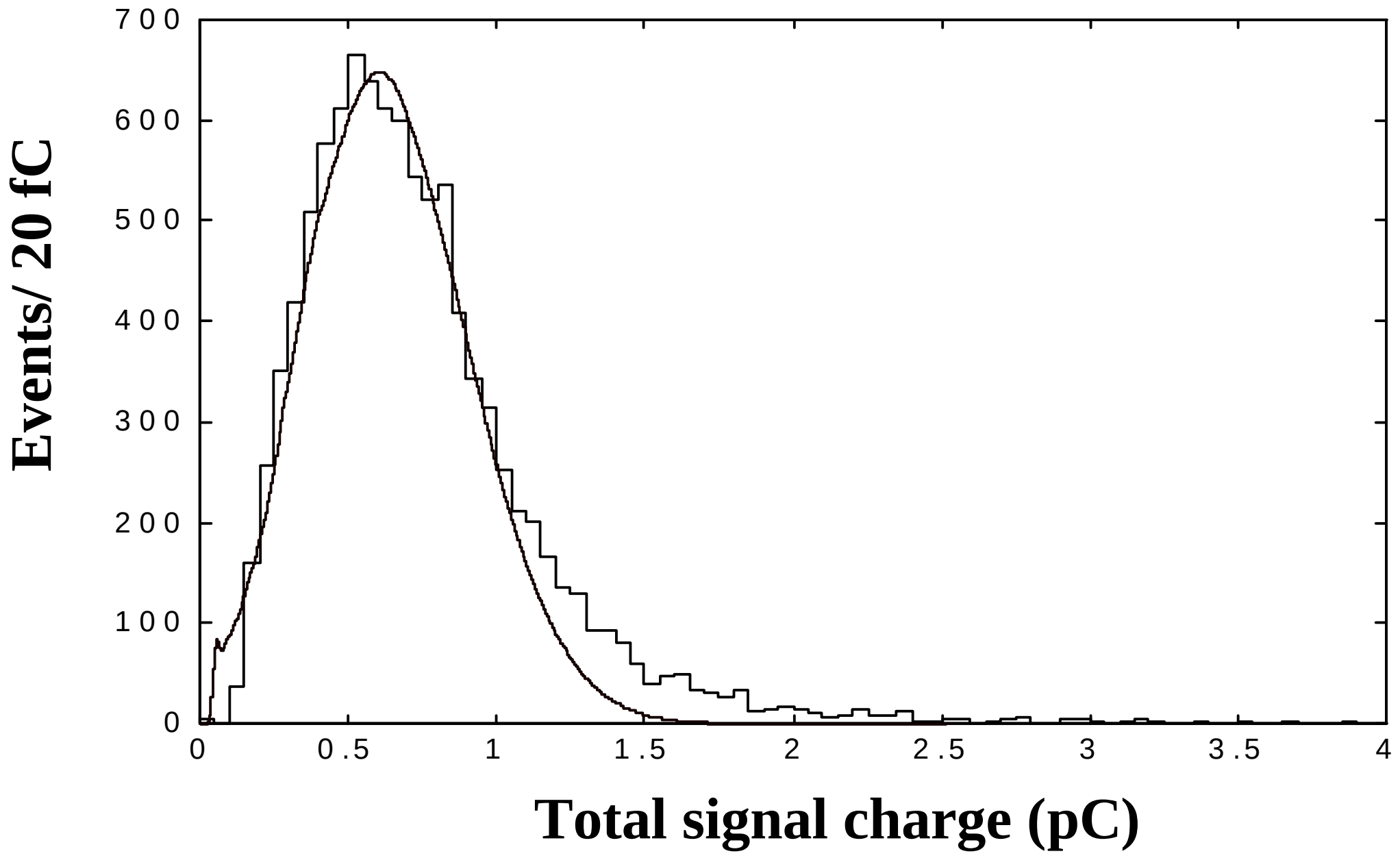

Figure 2 – Comparison between the signal charge distribution observed in a timing RPC (histogram) with four gas gaps and the 4-fold self-convolution of the charge distribution from a single-gap chamber (solid line) measured in similar operating conditions. The overall good agreement between both distributions suggests that the charge distribution observed in the four-gaps chamber can be interpreted as the analog sum of four independent single gaps.

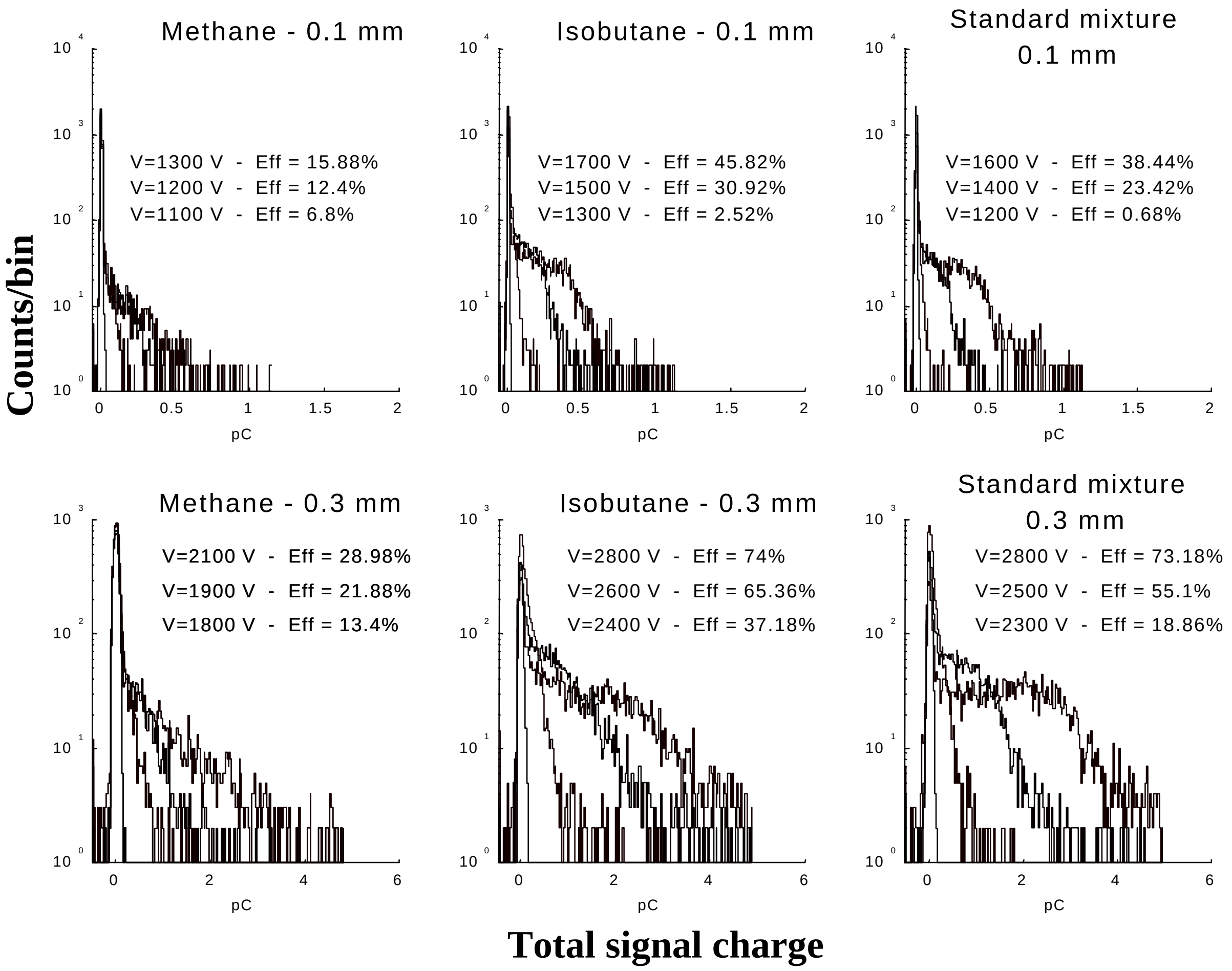

Figure 3 – Distribution of the signal charge in single gap chambers for several applied voltages, filling gases and gap widths. The innermost distributions correspond to the ADC pedestal and the peak close to 0 pC corresponds to the detector inefficiency. The efficiency figures were measured by the method described in [2].

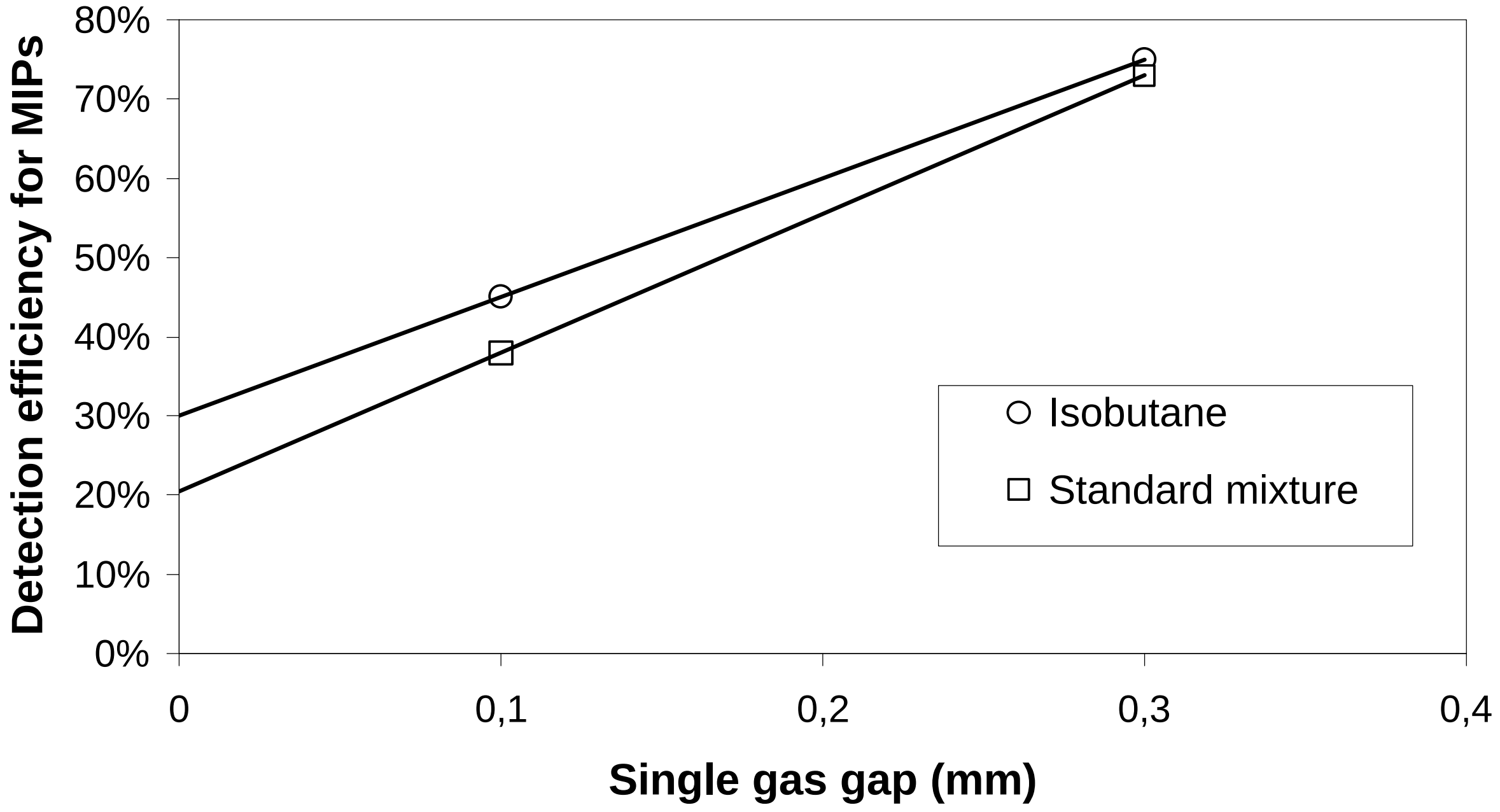

Figure 4 – The extrapolation of the observed detection efficiency to a vanishing gas gap width suggests that some additional process, other than gas ionization, may be contributing to the detection efficiency.

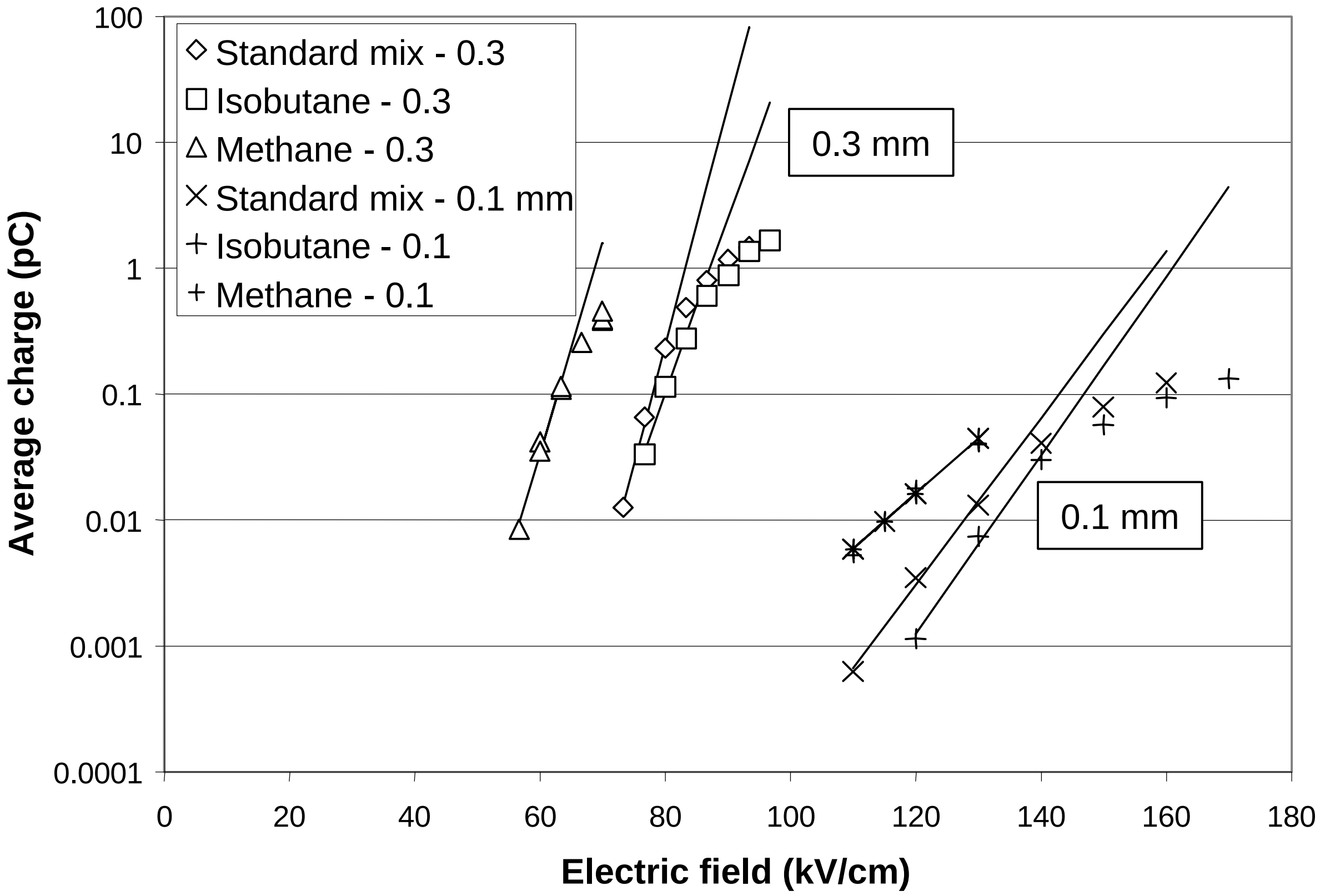

Figure 5 – Average signal charge as a function of the applied electrical field, calculated from the data presented in Figure 3. The solid lines correspond to exponential functions fitted to the lower 3 points of each experimental series and extrapolated to the larger fields, evidencing the sub-exponential character of experimental data (gain saturation). The onset of gain saturation for the 0.1 mm gaps occurs at a charge level that is an order of magnitude smaller than for the 0.3 mm gaps.

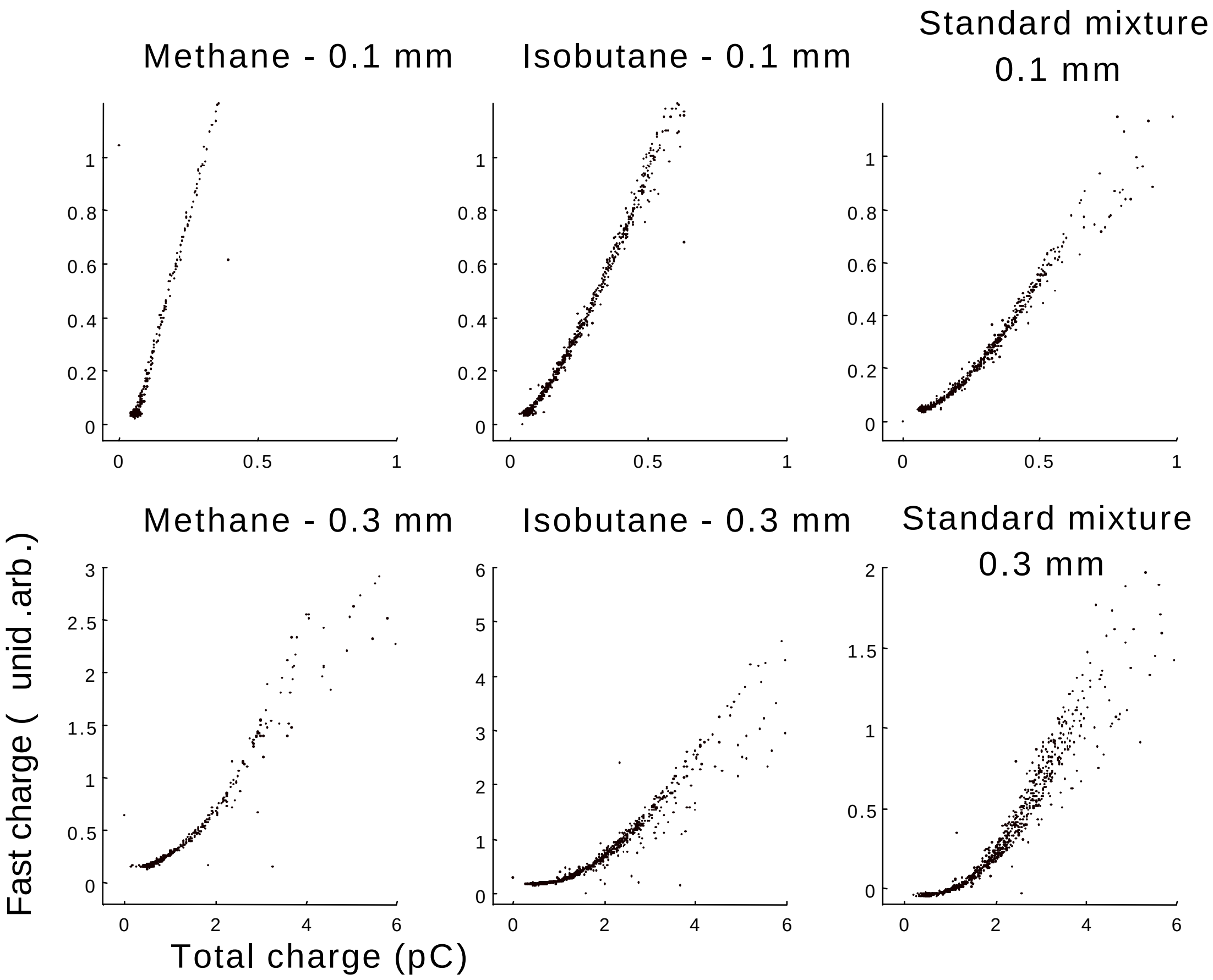

Figure 6 – Correlation plots between the fast (electron) signal charge and the total signal charge. Standard detector theory shows that the ratio between these quantities should be a constant, being the observed upward curving correlation compatible with a space charge effect.